\documentclass[manuscript]{aastex}
\usepackage{lscape}

\shorttitle{Helium Enhancement in the metal rich red giants of $\omega$ Centauri}
\shortauthors{Hema et al.}

\begin{document}

\title{Helium Enhancement in the metal rich red giants of $\omega$ Centauri}

\author{B. P. Hema\altaffilmark{1},  Gajendra Pandey\altaffilmark{1}, R. L. Kurucz\altaffilmark{2}, and C. Allende Prieto\altaffilmark{3}}

\affil{$^{1}$ Indian Institute of Astrophysics, Koramangala II Block, Bengaluru, Karnataka, India-560034 }

\affil{$^{2}$ Harvard-Smithsonian Center for Astrophysics, Cambridge, MA, 02138, USA}

\affil{$^{3}$ Instituto de Astrof\'{i}sica de Canarias, 38205 La Laguna, Tenerife, Spain ; Departamento de Astrof\'{i}sica, Universidad de La Laguna, 38206 La Laguna, Tenerife, Spain}

\begin{abstract}

The helium-enriched (He-enriched) 
metal-rich red giants of $\omega$ Centauri, discovered
by Hema \& Pandey using the low-resolution spectra from 
the Vainu Bappu Telescope (VBT) 
and confirmed by the analyses of the high-resolution
spectra obtained from the HRS-South African Large 
Telescope (SALT) for LEID 34225 and LEID 39048, 
are reanalysed here to determine their degree of 
He-enhancement/hydrogen-deficiency (H-deficiency). The 
observed MgH band combined with
model atmospheres with differing He/H ratios are used for the analyses.
The He/H ratios of these two giants are determined by enforcing the
fact that the derived Mg abundances from the Mg\,{\sc i} lines and 
from the subordinate lines of the MgH band must be same for the
adopted model atmosphere.
The estimated He/H ratios for LEID 34225 and LEID 39048 are
0.15$\pm$0.04 and 0.20$\pm$0.04, respectively, 
whereas the normal He/H ratio is 0.10.
Following the same criteria for the analyses of the 
other two comparison stars (LEID 61067 and LEID 32169), 
a normal He/H ratio of 0.10 is obtained. 
The He/H ratio of 0.15-0.20 
corresponds to a mass fraction of helium (Z(He)=Y) 
of about 0.375-0.445. The range of helium enhancement 
and the derived metallicity of the program stars are
in line with those determined for $\omega$ Cen's 
blue main-sequence stars. 
Hence, our study provides the missing link for the 
evolutionary track of the metal-rich helium-enhanced population 
of $\omega$ Centuari. 
This research work is the very first spectroscopic determination of
the amount of He-enhancement in the metal-rich red giants of 
$\omega$ Centauri using the Mg\,{\sc i} and MgH lines.

\end{abstract}


\section{Introduction}

$\omega$ Centauri, the largest and brightest 
milky way globular cluster is well known 
for hosting multiple stellar populations. The
cluster's metallicity is in the range 
$\sim$$-$2.5 $<$ [Fe/H] $<$ $\sim$$-$0.5 
\citep{norris96, suntzeff96, lee99, pancino00, sollima05,
johnson10}.

The survey was conducted to identify the 
hydrogen-deficient (H-deficient) stars in the globular cluster
$\omega$ Cen \citep{hema14} using the metal-rich ([Fe/H] $>$ $-$1)
giants of $\omega$ Cen.
The low-resolution spectra for the survey were obtained
from the Vainu Bappu Observatory in Kavalur, India.
The analysis was based on the strengths of the (0,0) MgH band and 
the Mg\,$b$ lines in the observed giant's spectrum. 
The Mg abundance derived from the MgH band was found to be 
two times or more lower than the average Mg abundance from 
Mg\,{\sc i} lines of $\omega$ Cen giants \citep{norris95}. 
This discrepancy in the Mg abundance from the MgH band and 
that from the Mg\,{\sc i} lines is attributed to the lower 
hydrogen abundance or helium enrichment, if not due to the 
uncertainties in the stellar parameters. 

The above study was confirmed by analysing the high-resolution
optical spectra of these giants obtained from the 
SALT observatory \citep{hema18}. Note that high-resolution spectra 
were obtained for two of the four H-deficient or He-enhanced 
giants discovered in our low-resolution spectroscopic 
survey \citep{hema14}. 
By deriving the Mg abundances from clean Mg\,{\sc i} lines
and from the subordinate lines of the (0,0) MgH band, 
\citet{hema18} confirm 
that there a discrepancy exists in the Mg abundances derived 
from these two indicators. This discrepancy, if not due to 
the uncertainties in the star's derived 
effective temparature and surface gravity, is then due to 
the less hydrogen or more helium in the star's atmosphere. 

In this paper we present the procedure for estimating the amount of
H-deficiency/He-enhancement in a star's atmosphere by adopting 
model atmospheres with normal and differing 
He/H ratios. The observations, 
the abundance analysis procedure, and the adopted model atmospheres
are described in the following sections. The results 
are discussed in the
light of He-enrichment in $\omega$ Cen.

\section{Observations}

The high-resolution optical spectra were 
obtained using Southern African Large
Telescope (SALT) $-$ high resolution spectrograph
(HRS)\footnote{SALT HRS is a dual beam fibre-fed,
white-pupil, echelle spectrograph, employing
VPH gratings as cross dispersers.}.
These spectra obtained with the SALT-HRS have
a resolving power, R ($\lambda$/$\Delta$$\lambda$) of
40000. The spectra were obtained with both the blue and
red cameras using 2K $\times$ 4K and 4K $\times$ 4K CCDs,
respectively, spanning a spectral range of
370$-$550 nm in the blue and 550$-$890 nm in the red.

The spectral reductions were carried out using the IRAF
\footnote{The IRAF software is distributed by the
National Optical Astronomy Observatories under contract
with the National Science Foundation.}
(Image Reduction and Analysis Facility)
software. The traditional data reduction procedure, including
bias subtraction, flat-field correction, spectrum extraction,
wavelength calibration, etc., was followed.
The extracted and wavelength-calibrated 1D spectra were
continuum normalized. The region of the spectrum with
maximum flux points and free of absorption lines was
considered for continuum fitting with a smooth curve passing
through these points. 
Each of our program star was observed for about 30 minutes exposure 
in red and blue camera. To improve the signal-to-noise ratio, the
observed spectra of the program stars were smoothed such that
the strengths of the spectral lines are not altered. The signal-to-noise
ratio per pixel, in the continuum, is ∼150 for the smoothed 
blue spectra of
the program stars at about 5000\AA\ and $\sim$200 for red spectra at
about 7000\AA. Since there is an overlap of wavelengths, the
spectrum is continuous without gaps in the blue as well as in the red
spectral range. The atlas of high-resolution spectrum of Arcturus
(Hinkle et al. 2000) was used as a reference for continuum
fitting and also for identifying the spectral lines \citep{hema18}.

\section{Abundance Analysis}

The processed observed spectrum, as explained above, was used 
for conducting the abundance analysis. 

The equivalent widths for several clean lines, weak and strong, 
were measured using the tasks 
in the IRAF software package.
Using the measured equivalent widths, and the 
LTE line analysis and spectrum synthesis code MOOG 
\citep{sneden73} combined with model atmospheres, 
the stellar parameters: effective temperature, surface gravity, and
microturbulence, were determined.
The ATLAS9 \citep{kurucz98} plane parallel, line-blanketed
LTE model atmospheres were adopted for the analysis.
The derived abundances using MOOG are based on the adopted 
model atmosphere's He/H ratio. The input abundances of 
H and He provided to MOOG, that adopts a model atmosphere
computed for a normal He/H ratio of 0.1, are log $\epsilon$ (H)=12 and
log $\epsilon$ (He)=11, respectively. Nevertheless, the input abundances
of H and He provided to MOOG are log $\epsilon$ (H)=11.894 and 
log $\epsilon$ (He)=11.195, respectively, if a model atmosphere
of He/H ratio 0.2 is adopted for analysis.
The abundances for H and He were as usual calculated for different
He/H ratios, assuming H and He as the major
constituents of the stellar composition and all other elements
are only in trace amounts. For example, if He/H ratio of 0.2 is adopted,
then log $\epsilon$ (H)=11.894 and log $\epsilon$ (He)=11.195 are
obtained using the standard 
relation: $n_{\rm H}$+ 4 $n_{\rm He}$=10$^{12.15}$.
Similarly, for He/H ratio of 0.15, log $\epsilon$ (H)=11.945 and 
log $\epsilon$ (He)=11.121 are obtained.

The procedure for determining the stellar parameters 
for the program stars is described and executed in \citet{hema18}.
The complete linelist used for the abundance analysis and 
the line-by-line abundances are given in TABLE 2 of 
\citet{hema18}. Hema et al's determination of stellar parameters 
for the analysed stars are in excellent agreement with 
those derived by \citet{johnson10}.
Hence, the stellar parameters for the current study have been 
adopted from \citet{hema18}.

\subsection{Mg abundance: MgH band and the Mg\,{\sc i} lines}

Our region of interest is the blue degraded (0,0) MgH band 
extending from 5330 to 4950 \AA\ with the band head at 5211\AA\
and the Mg\,$b$ lines at 5167.32\AA, 5172.68\AA, and 5183.60\AA.
From \citet{hema14, hema18} study, based on the strengths of the 
MgH band and the Mg\,$b$ lines in the observed low-resolution spectra, 
four stars were identified with having strong 
Mg\,$b$ lines but weaker/weakest (0,0)
MgH band than expected for their stellar parameters. The low-resolution 
spectra were obtained from the Vainu Bappu Telescope in Kavalur, 
India. The stars were divided into three groups based on the 
strengths of the Mg\,$b$ lines and the MgH band. The first group 
corresponds to metal rich stars having strong Mg\,$b$ lines and strong 
MgH band, the second group include metal poor stars having 
weak Mg\,$b$ lines and a weak MgH band, and the third group are metal 
rich stars having strong Mg\,$b$ lines with a weak MgH 
band, in their observed low-resolution spectra. The stellar 
parameters for the program stars were derived using  
photometric colours (Johnson and Str\"{o}mgren) 
by \citet{hema14} which 
were also in good agreement with those derived spectroscopically 
by \citet{johnson10}. The spectra were analysed using the 
spectrum synthesis code $synth$ in MOOG. 
The analysis was carried out by 
comparing the Mg abundances derived from the subordinate lines
of the MgH band with that of the average Mg abundance derived 
for the red giants of $\omega$ Cen using Mg\,{\sc i} lines 
\citep{norris95}. 
The Mg abundances derived from the MgH band are expected to 
be same with that
derived from Mg\,{\sc i} lines. If there is a
discrepancy in the derived Mg abundances, and if the difference 
is not due to the uncertainties in the stellar parameters, then  
the observed MgH band weaker than expected is attributed to the 
atmosphere's lower hydrogen abundance than normal.

Two metal rich stars from the first group (LEID 39048 and 
LEID 60073) and two metal rich stars from the third group 
(LEID 34225 and LEID 35201) were identified with having stronger 
Mg\,$b$ lines and weaker MgH band than expected for their effective 
temperatures \citep{hema14, hema18}. 
For one of the first group (LEID 34225) and one from the third 
group (LEID 39048), along with their comparison stars (LEID 61067
and LEID 32169), high-resolution
spectra from the SALT observatory were obtained.
Using the spectrum synthesis technique, described in Section 3 of 
\citet{hema14}, the spectra were analyzed.
A detailed abundance analysis has been carried out by \citet{hema18}
for the two program stars (LEID 34225 and LEID 39048)
along with their comparison stars (LEID 61067 and LEID 32169).
Using the measured equivalent widths, the stellar parameters 
and the elemental abundances were derived.
For these four stars (two program and their two comparison stars), 
the Mg abundance from the weaker Mg\,{\sc i} lines were derived and
contrasted with that derived from the subordinate lines of the MgH band.

The Mg abundance derived for two comparison stars, which 
were identified as normal \citet{hema14}, have the 
same Mg abundance from the Mg\,{\sc i} lines as well as from the 
weaker subordinate lines of the MgH band within the uncertainties.
But, for the two identified H-poor stars 
(here program stars: LEID 34225 and LEID 39048), the 
difference in the Mg abundance from the Mg\,{\sc i} and that from 
the MgH band is greater than 0.3 dex (see Table 6 of \citet{hema18}). 
All sources of uncertainties were explored and confirmed that this 
discrepancy is not reconcilable by changing the stellar parameters within 
the uncertainties. Since the Mg abundance for the program stars is fixed 
from the weak Mg\,{\sc i} lines, this difference cannot be attributed to 
the Mg abundance forming the MgH band but to the hydrogen abundance. 

Hence, in this paper, model atmospheres with differing He/H ratio 
than normal are used to determine the degree of He-enrichment 
or H-deficiency in the star's atmosphere.

\subsection{Model atmospheres}

The observed spectra of the program stars were reanalysed 
using model atmospheres with differing He/H ratios  
computed by one of us (RLK). 

     We produced one-dimensional ATLAS12 
opacity-sampling model atmospheres for the range of effective 
temperatures and gravities, scaled-solar metal abundances, and 
the helium abundances appropriate for this analysis.
ATLAS12 is described in \citet{kurucz14}. The atomic and molecular 
line lists were described in \citet{kurucz17}
and again in \citet{kurucz18}.

In addition to this, we wrote 
scripts\footnote{Model atmospheres with varying He/H 
ratios were computed using ATLAS9 (\citet{kurucz93} 
and subsequent updates), in particular the version of 
the code set up to compile and run in gnu-linux by 
\citet{sbordone05, sbordone04, sbordone07},
(http://atmos.obspm.fr/). The  line opacities used 
are those provided in the ODFs distributed with the code 
(tagged as NEWODF; \citet{castelli03}, and the solar reference 
metal mixture is that of \citet{grevesse98}. For convenience 
the code was run using a  Perl wrapper that sets up the relevant 
input/output files and iterates until the models converge.} 
that iterate and generate Kurucz LTE plane 
parallel model atmospheres with a fine grid in He/H ratios from
the above discussed coarse grid of model atmospheres.

\subsection{Determination of the He/H ratio}

The stellar parameters, effective temperature=$T_{\rm eff}$, 
surface gravity=$\log g$ and microturbulence=$\xi$, 
for the program stars are from \citet{hema18}. The adopted 
model atmospheres by \citet{hema18} for deriving stellar 
parameters are of normal He/H ratio 0.10. The stellar parameters 
were rederived using a grid of model atmospheres with
He/H=0.15, 0.20, 0.25, and 0.30; the adopted metallicity 
of the grid is fixed based on the derived 
metallicity of the program star from \citet{hema18}. 
The procedure adopted for 
determining the stellar parameters is ditto as described in 
Section 3 of \citet{hema18}. The rederived stellar parameters 
are not sensitive and almost independent of the 
adopted grid's He/H ratio: 0.10, 0.15, 0.20, 0.25, and 0.30.

For the program star's derived stellar parameters, synthetic spectra
in the MgH band region were computed for different He/H ratios. 
The Mg abundances used for the above 
syntheses were derived from the measured equivalent 
widths of the weak Mg\,{\sc i} 
lines for the adopted model's He/H ratio. Hence, the best fit 
to the MgH band in the observed spectrum determines the 
adopted model's He/H ratio.
Finally, the elemental abundances are derived for
the adopted stellar parameters: $T_{\rm eff}$,
$\log g$, $\xi$, and the He/H ratio. Note that
the adopted metallicty of the model atmosphere comes from
the iron abundances derived from the measured equivalent widths of 
the Fe lines for the star in question, and is an iterative process.
The results of our analyses with adoption of $\alpha$-enhanced 
model atmospheres having [O/Fe]=0.5, 
are in excellent agreement with that of solar scaled model 
atmospheres having [O/Fe]=0.

In the case of elemental abundances derived 
by adopting model atmospheres
with same stellar parameters but varying He/H ratios (for example,
He/H=0.20 and 0.10), the abundances from the higher
He/H ratios are lower than that derived from the normal.
Here, decreasing the hydrogen abundance or increasing the 
helium abundance i.e., increasing the He/H ratio, 
lowers the continuous opacity per gram \citep{sumangala11}. Hence, 
for the same observed strength of the spectral line, 
the elemental abundance has to decrease (see Table 1).
In Table 1, for the program stars, the abundances 
derived for the normal He/H ratio of
0.10 and also for the determined He/H ratio are given. 
The Mg\,{\sc i} line at $\lambda\lambda$5711\AA\ is synthesized 
for the program stars LEID 34225 and LEID 39048 (see Figure 1).
The determination of the Mg abundance from the 
Mg\,{\sc i} lines by equivalent width analysis is in 
good agreement with that of the synthesis of 
$\lambda\lambda$5711\AA.
The decrease in abundance is proportional to the amount 
of H-deficiency or the He-enrichment applied in the sense of
He/H ratio. 
However, the abundance ratios remain unchanged for most 
of the elements with few exceptions.

Examples of the spectrum synthesis in the MgH band region for
the program stars are presented in figure 2 and 3.
By adopting the respective elemental abundances derived from
the adopted model computed for the pair of [Fe/H] and the He/H ratio, 
spectra in the MgH band region were synthesized. 
Figure 2, for the program star LEID 34225, clearly shows that
for the Mg abundance of 6.52$\pm$0.02 (from Mg\,{\sc i} lines), the 
best fit to the observed MgH band (mainly the subordinate 
lines of MgH band at about 5175\AA) is obtained for He/H 
ratio of 0.15. 
Similarly, for LEID 39048, with the Mg\,{\sc i} abundance 
of 7.25$\pm$0.06, the best fit to the observed MgH band is 
obtained for He/H ratio of 0.20 (see Figure 3).

The He/H ratio for the program stars is determined from 
the observed MgH band in their spectra. Once the stellar 
parameters and metallicity are determined and fixed, 
the MgH band strength depends mainly on the 
atmosphere's Mg abundance and the He/H ratio.
Then, the uncertainty on the He/H ratio is primarily due to 
the uncertainty on the Mg abundance. Note that, the uncertainty 
in the Mg abundance for the program star LEID 34225 is about 
0.02 dex and that for LEID 39048 is about 0.06 dex. 
In general, the uncertainty on the He/H ratio is about 
0.04 dex.

The corresponding mass fraction for the derived He/H ratio 
for the program stars: 
LEID 34225, He/H\,=\,0.15$\pm$0.04, 
Z(H)=0.625 and Z(He)=Y=0.374; 
LEID 39048, He/H\,=\,0.20$\pm$0.04, 
Z(H)=0.555, Z(He)=Y=0.445.

For the two normal comparison stars from our previous
study \citet{hema18}, the Mg abundance derived 
from the Mg\,{\sc i} lines is same as that from the
MgH band within the errors. Hence, these stars 
have the normal value of He/H\,=\,0.10.

\section{Results and Discussion}

The multiple stellar populations are present in 
almost all the studied Galactic globular clusters (GGC) 
($\sim$ 60), including $\omega$ Centauri, and also in several 
extragalactic globular clusters \citep{milone18, 
marino12, marino09, marino2011a, 
yong08, milone20}.
The red giant branch (RGB) of $\omega$ Centauri show a large spread in 
metallicity from the mean metallicity of the cluster,
[Fe/H]: $\sim$$-$2.5 $<$ [Fe/H] $<$ $\sim$$-$0.5 \citep{bedin04,
sollima05, johnson10, Simpson13}. 
Along with the complex stellar populations in these globular 
clusters, they also show helium enhancement
among the main-sequence stars, the red giants and also 
the horizontal branch (HB) stars \citep{milone18}. 
The He-enhancement in the main-sequence stars of 
$\omega$ Centauri have been studied
by \citet{piotto05}. Similarly, a few clusters for example,
NGC 2808 \citep{piotto07}, NGC 6752 and NGC 6397 
\citep{Milone10, milone12, milone13}, 
in which the main-sequence splits
due to He-enhancement have been studied.
The He-enhancement among the blue horizontal
branch stars have also been studied for 
four of NGC 6752 and six in M4 \citep{villanova09, marino11b}.

From the studies of \citet{piotto05} and \citet{milone17}, there are two
main-sequence branches identified among the main
sequence populations in $\omega$ Centauri. They are
the $red$ main-sequence (rMS) and the $blue$ main-sequence
(bMS). From the spectroscopic studies of rMS and bMS stars, Piotto et al.
have determined that the rMS stars are more metal-poor ([M/H]=$-$1.57)
than the bMS stars ([M/H]=$-$1.26).
According to the canonical stellar models
with canonical chemical abundances, the bMS should be
more metal-poor than the rMS. To account for the fact that
the bMS stars are less metal-poor, the only explanation could
be the helium-enhancement in their atmospheres. For their derived
metallicities, the isochrones calculated using the models with different
helium abundances were compared with the observed CMD of
$\omega$ Cen (see Figure 7 of \citet{piotto05}).
The bMS can be reproduced only by assuming Y $>$ 0.35, that 
corresponds to He/H $>$ 0.13. 
From the near infrared transition of He\,{\sc i} at 1.08$\mu$m, 
the direct measure of He-abundance has been obtained by 
\citet{dupree11, dupree13} for the red giants of $\omega$ Cen . 
These giants are metal poor having the mean metallicity of the cluster
([Fe/H]$\sim-$1.7). 
This provides a clue that the clusters with multiple stellar 
population may also host the He-enhanced population. The  
helium content measured in these 
red giants is about Y$\leq$0.22 for one and Y=0.39-0.44 
for another which is similar to that measured for the main-sequence 
stars of $\omega$ Cen.

From our study of the metal rich red giants of $\omega$ Cen, as discussed 
in the Sections above, the He/H ratio is determined based on 
no difference in the Mg abundance derived from
the Mg\,{\sc i} lines and that from the MgH band.
A proper He/H ratio stellar model is 
identified such that the Mg abundance from the MgH band 
matches with that from the Mg\,{\sc i} lines. 
From the analyses of the program stars, 
the He/H ratio estimated are
0.15$\pm$0.04 and 0.20$\pm$0.04 for LEID 34225 and LEID 39048,
respectively. These correspond to a mass fraction for
LEID 34225: He/H\,=\,0.15$\pm$0.04, Z(H)=0.625 and 
Z(He)=(Y)=0.374, and for
LEID 39048: He/H\,=\,0.20, Z(H)=0.555, Z(He)=(Y)=0.445.
For calculating the mass fraction, only hydrogen and helium
are considered, as the other elements are in trace amounts.

As discussed in our previous analysis by \citet{hema18},
our helium rich stars are similar to the bMS stars, having
similar metallicities. Out of four
program stars, two were helium-rich (LEID 34225 and
LEID 39048) and two were
normal (comparison stars: LEID 61067 and LEID 32169).
In this paper, using the models with appropriate He/H ratio,
the metallicities for the two helium rich program stars are determined.

Using the apropriate He/H ratio derived by the
MgH band, the derived metallicities for these He-rich
program stars are,
for LEID\,34225, He/H=0.15 (Y=0.375), [Fe/H]=$-$1.2, and
for LEID\,39048, He/H=0.20 (Y=0.445), [Fe/H]=$-$0.8.
Hence, the newly derived metallicities, are in excellent agreement
with the metallicities of the bMS stars ([Fe/H=$-$1.26)
within the uncertainties.
And, the He content from our studies for giants of
$\omega$ Cen is about Y=0.375 and 0.445.
These values are same as that observed for the
bMS stars. This is a strong indication that, these He-enhanced
stars share the same evolutionary connections.
There is a metal rich population present from the MS stars 
through the horizontal branch stars of $\omega$ Cen \citep{piotto05,
lee99, pancino00, johnson10, Villanova07, cassisi09, dantona10,
bellini10}. Among these, He-enhancement is spectroscopically 
confirmed in the blue-MS \citep{piotto05} and the red-giants of our 
sample \citep{hema14,hema18}.
Presence of the multiple branches in different evolutionary 
stages is a strong clue for the presence of the He-enhanced population.
The direct measure of helium in the He-enhanced population are 
also been studied in the other globular clusters of the Galaxy such as
NGC 2808 \citep{marino14}.

$\omega$ Cen exhibits a unique and more complex
chemical  enrichment. Some of the possible scenarios 
that could account for this enrichment would be:
 multiple star-formation episodes, contribution from 
the massive stars which would end up as Type II SNe, 
the mass loss by the asymptotic giant branch stars, 
or may be all these scenarios have played a role in producing the
observed abundance pattern of MS, subgiant branch (SGB), RGB and HB stars. 
The observed anomalies such as the spread in the 
metallicity in the MS, SGB, RGB and HB, and also the
enhancement in $s$-process elements would favour the 
occurence of the multiple star-formation episodes. 
The pattern of chemical enrichment observed in the 
metal poor red giants may be due to the rapid evolution 
of massive stars which ended up as Type II SNe. 
The significant contribution for the enhancement of Na, 
Al and s-process elements may come from the mass loss 
by the RGB and AGB stars. 
The multiple star-formation episodes taking place 
at different intervals might have formed the 
stars with different metallicities with different stellar 
composition \citep{johnson10, milone18, marino12}.

Our discovery of He-enhanced
red giants with determination of the helium-enhancement and
their metallicity, is same as that of the bMS stars.
Hence, bMS stars are most probably the progenitors
of the metal rich subgiants, red giants and horizontal branch 
stars of $\omega$ Cen. Though, there are metal-poor RGB stars that are 
He-enhanced, the bMS stars may be the progenitors for the 
metal-rich He-enhanced population indicating that, these
are formed at the same epoch from the same material. Hence, this also 
rules out the fact that only the metal-rich 
giants are He-enehanced. However, more studies are required 
to investigate the He-enhancement in the cluster stars with 
different metallicities in the SGB, RGB and HB. For cool stars,
the technique of contrasting the Mg abundances from the Mg\,{\sc i} 
lines and from the subordinate lines of the (0,0) MgH band 
is very promising and a novel method to investigate the 
H-deficiency/He-enhancement. Our studies \citep{hema14, hema18} 
are the first spectroscopic studies to investigate, discover and 
confirm the  He-enhancement in the metal-rich giants 
of $\omega$ Centauri, providing a crucial information and a 
link for understanding the evolution of metal-rich He-enhanced stars.

\section{Conclusions}

From the discovery of the He-rich red-giants in the
Galactic globular cluster $\omega$ Centauri using the
low-resolution spectra \citep{hema14},
study of these giants by obtaining the high-resolution spectra
for deriving the appropriate Mg abundance from the Mg\,{\sc i}
to confirm the H-deficiency/He-enrichment using the MgH bands
\citep{hema18}, and now estimating the H-deficiency/He-enrichment
using the proper model atmopsheres, confirms the existence of
He-enhanced metal-rich red giants same as that observed
for the blue main-sequence stars of $\omega$ Centauri.
Our studies bridges the evolutionary track of the metal-rich 
He-rich population of $\omega$ Cen.

\section{Acknowledgment}

We thank the anonymous referee for a constructive report.


\clearpage

\begin{figure}
\plotone{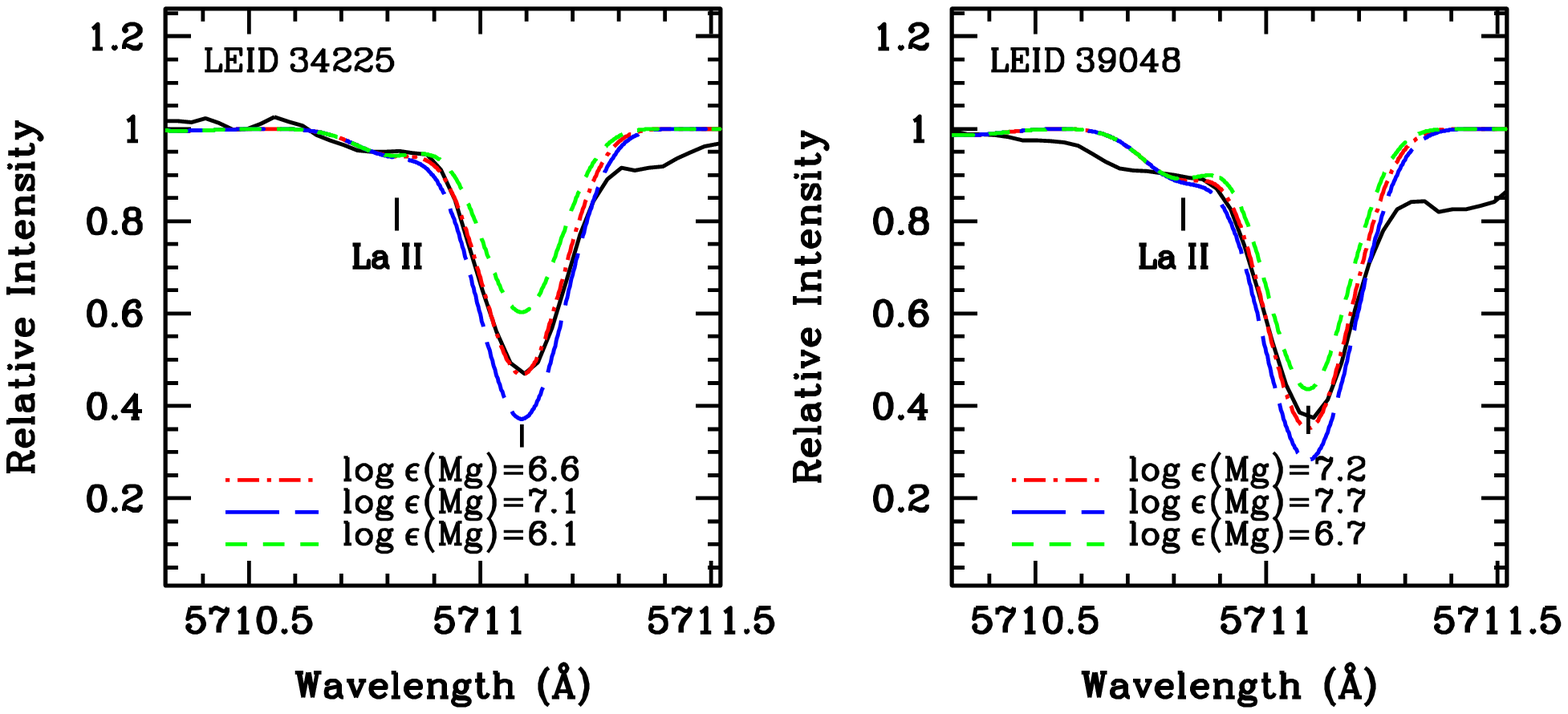}
\caption{The Mg abundance derived from the $\lambda\lambda$5711\AA\
Mg\,{\sc i} line for the program stars. 
The best fit synthesis is shown with the red-dash dotted line.}
\end{figure}

\begin{figure}
\plotone{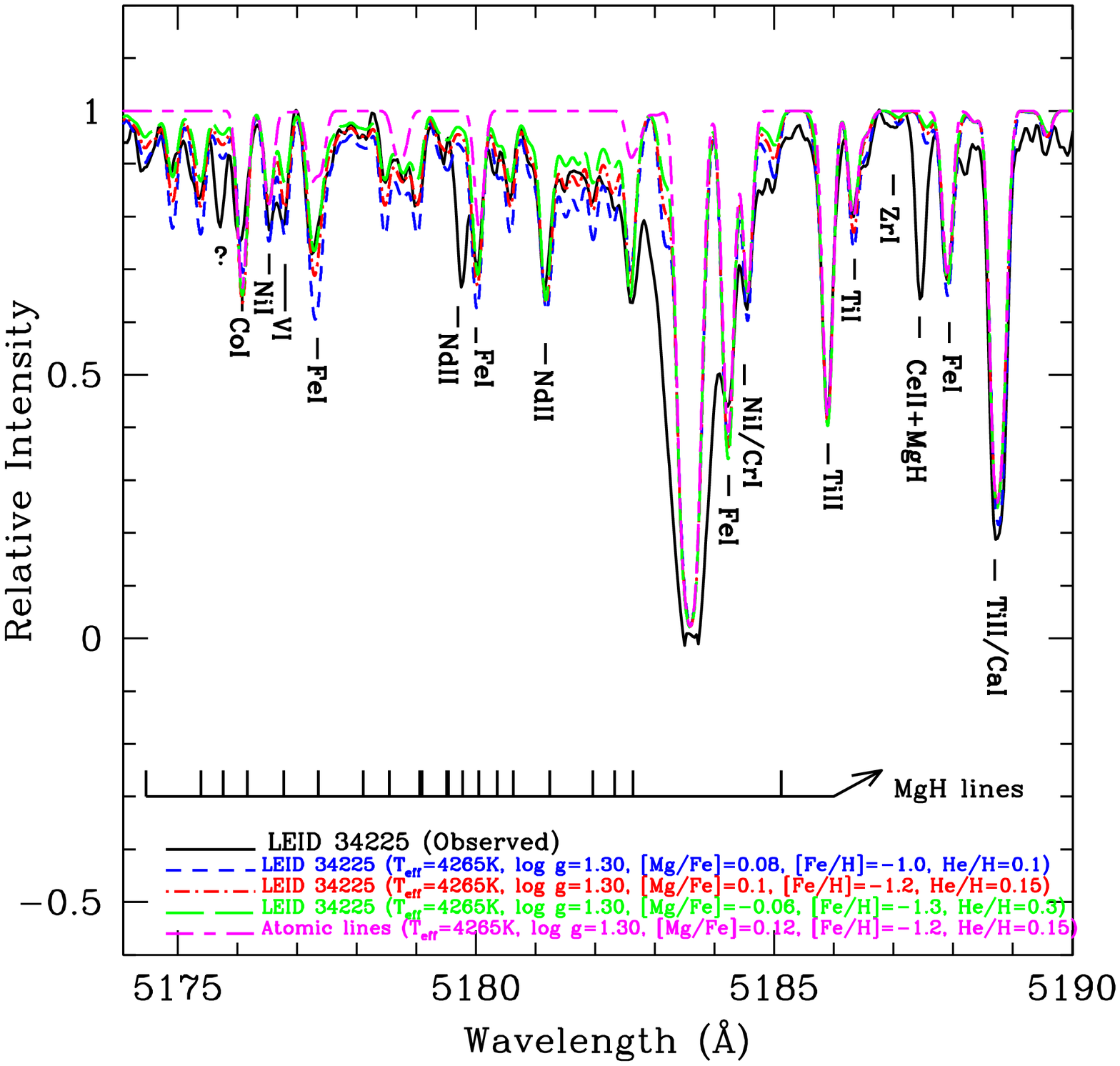}
\caption{Observed and the synthesized MgH bands for LEID 34225 are shown.
The spectra synthesized for the Mg abundance derived from the 
Mg\,{\sc i} lines and the best fit value of He/H ratio are 
shown in red dash-dotted line. The synthesis for the two 
value of the He/H are also shown. }
\end{figure}

\clearpage

\begin{figure}
\plotone{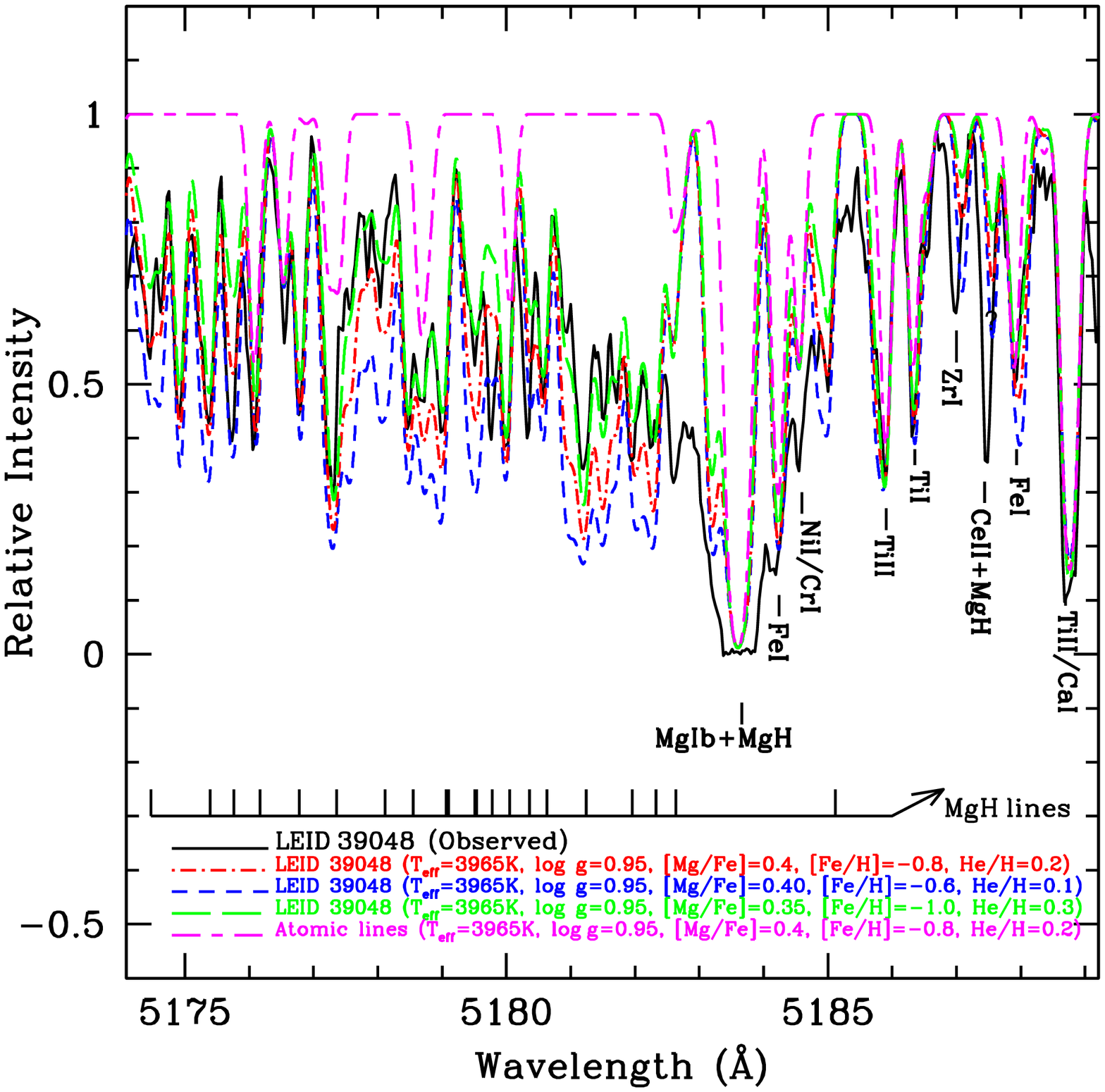}
\caption{Observed and the synthesized MgH bands for LEID 39048 are shown.
The spectra synthesized for the Mg abundance derived from the
Mg\,{\sc i} lines and the best fit value of He/H ratio are
shown in red dash-dotted line. The synthesis for the two
value of the He/H are also shown.}
\end{figure}

\clearpage

\thispagestyle{empty}
\begin{landscape}
\begin{deluxetable}{rrrrrrrrrrrrrrr}
\tabletypesize{\tiny}
\tablecolumns{13}
\tablewidth{0pc}
\tablecaption{Abundances for different He/H ratios}
\tablehead{
\colhead{} & \colhead{}   &  \multicolumn{5}{c}{LEID 34225} &   \colhead{}   &
\multicolumn{5}{c}{LEID 39048} \\
\cline{3-7} \cline{9-13} \\
\colhead{Elements} & \colhead{log $\epsilon\odot$}   & \colhead{log $\epsilon$(He/H=0.1)}    & \colhead{[X/Fe]} &
\colhead{log $\epsilon$(He/H=0.15)}    & \colhead{[X/Fe]}   & \colhead{$n$}  & \colhead{}  & \colhead{log $\epsilon$(He/H=0.1)} & \colhead{[X/Fe]} & \colhead{log $\epsilon$(He/H=0.2)} & \colhead{[X/Fe]} & \colhead{$n$}}
\startdata
H & 12.00 & 12.00 & \nodata & 11.945 & \nodata  &\nodata &&  
12.00 & \nodata  & 11.894 & \nodata & \nodata \\
He & 10.93 & 11.00 &  \nodata & 11.121 & \nodata  &\nodata && 
11.00 & \nodata & 11.195 & \nodata & \nodata \\
O & 8.69 & 7.4 & -0.25 & 7.36 & -0.13 & 1 && 
8.1$\pm$0.06 &  0.03 & 7.99$\pm$0.07 & 0.10 & 2 \\ 
Na & 6.24 & 5.72$\pm$0.16 & 0.52 & 5.65$\pm$0.16 & 0.61 & 4 &&  
6.50$\pm$0.04 & 0.88 & 6.38$\pm$0.04 & 0.94 &2 \\ 
Mg (Mg\,{\sc i}) & 7.60 & 6.65$\pm$0.05 & 0.09 & 6.52$\pm$0.02 & 0.12 & 4 && 
7.41$\pm$0.1 & 0.43 & 7.25$\pm$0.06 & 0.45 & 5 \\
Mg (MgH) & \nodata &  6.26 & $-$0.29 & 6.50 & 0.10 & \nodata & 
& 7.0 & 0.02 & 7.20 & 0.40 & \nodata \\
Al & 6.45 & 6.50$\pm$0.12 & 1.09 & 6.32$\pm$0.11 & 1.07 & 4 && 
6.50$\pm$0.13 & 0.67 & 6.33$\pm$0.1 & 0.68 & 4\\
Si & 7.51 & 7.00$\pm$0.11 & 0.53 & 6.85$\pm$0.11  & 0.54 & 5 && 
7.36$\pm$0.15 & 0.47 & 7.02$\pm$0.16 & 0.31 & 7 \\
Ca & 6.34 & 5.40$\pm$0.16 & 0.1 & 5.26$\pm$0.16 & 0.12 & 9 && 
5.9$\pm$0.08 & 0.18 & 5.82$\pm$0.08 & 0.28 & 8 \\
Sc\,{\sc i} & 3.15 & 2.24$\pm$0.16 & 0.13 & 2.20$\pm$0.14 & 0.25 & 4 && 
\nodata & \nodata &  \nodata & \nodata & \nodata \\
Sc\,{\sc ii} & 3.15 & 2.25$\pm$0.09 & 0.14 & 2.13$\pm$0.09 & 0.18 & 5 && 
2.60$\pm$0.2 & 0.07 & 2.37$\pm$0.20 & 0.02 & 5 \\
Ti\,{\sc i} & 4.95 & 4.30$\pm$0.14 & 0.39 & 4.24$\pm$0.14 & 0.49 & 7 && 
4.80$\pm$0.17 & 0.47 & 4.80$\pm$0.17 & 0.67 & 11 \\
Ti\,{\sc ii} & 4.95 & 4.29$\pm$0.11 & 0.38 & 4.14$\pm$0.11 & 0.39 & 4 && 
\nodata & \nodata & \nodata & \nodata & \nodata \\
V & 3.93  & 2.66$\pm$0.14 & -0.23 & 2.64$\pm$0.14 & -0.09 & 6 && 
3.20$\pm$0.15 & -0.11 & 3.20$\pm$0.15 & 0.08 & 8 \\
Cr & 5.64 & 4.60$\pm$0.13 & 0.0 & 4.53$\pm$0.13  & 0.09 & 7 && 
5.00$\pm$0.1 & -0.02 & 4.92$\pm$0.11 & 0.08 & \nodata \\
Mn & 5.43 & 4.39$\pm$0.06 & 0.0 & 4.32$\pm$0.07 & 0.09 & 3 &&  
4.85$\pm$0.04 & 0.04 & 4.77$\pm$0.04 & 0.14 & 3 \\
Fe\,{\sc i} & 7.50 & 6.46$\pm$0.16 & -1.04 & 6.32$\pm$0.12 & -1.18 & 19 && 
6.88$\pm$0.14 & $-$0.62 & 6.70$\pm$0.13 & $-$0.80 & 16 \\
Fe\,{\sc ii} & \nodata & 6.46$\pm$0.06 & -1.04 & 6.25$\pm$0.08 & -1.25 & 2 && 
\nodata & \nodata & \nodata & \nodata & \nodata \\ 
Co & 4.99 & 4.01$\pm$0.14 & 0.06 & 3.92$\pm$0.14 & 0.13 & 6 && 
4.32$\pm$0.13 & -0.05 & 4.14$\pm$0.11 & -0.05 & 7\\
Ni & 6.22 & 5.13$\pm$0.09 & -0.05 & 5.04$\pm$0.09 & 0.02 & 3  && 
5.74$\pm$0.17 & 0.14 & 5.50$\pm$0.18 & 0.09 & 7 \\
La & 1.10 & 1.10 & 1.04 &  0.95 & 1.05 & 1 && 
1.32 & 0.84 & 1.19 & 0.89 & 1 \\ 
\enddata
\end{deluxetable}
\end{landscape}

\end{document}